\documentclass{svproc}
\usepackage{graphicx}%
\usepackage{multirow}%
\usepackage{amsmath,amssymb,amsfonts}%
\usepackage{mathrsfs}%
\usepackage[title]{appendix}%
\usepackage{xcolor}%
\usepackage{textcomp}%
\usepackage{manyfoot}%
\usepackage{booktabs}%
\usepackage{algorithm}%
\usepackage{algorithmicx}%
\usepackage{algpseudocode}%
\usepackage{listings}%
\usepackage{subcaption}
\usepackage{hyperref}
\usepackage{orcidlink}

\usepackage{url}

\begin{document}
\mainmatter              
\title{Extracting Illicit-Rich Subgraphs from Transaction Networks via Feed-Guided Shortest Paths}
\titlerunning{Strategic Feed Selection for SPNSA in AML Graphs}  
%
\author{%
Doniyor Koshimbetov\inst{1}\orcidlink{0009-0002-6434-5959} \and
Emine \c{S}ule Yaz\i c\i\inst{1}\orcidlink{0000-0001-6824-451X} \and
Asha Rao\inst{2}\orcidlink{0000-0001-6222-282X} %
}
\authorrunning{Doniyor et al.}

\institute{%
Department of Mathematics, College of Sciences, \\ Ko\c{c} University, \.{I}stanbul 34450, T\"urkiye\\
\email{dkoshimbetov22@ku.edu.tr, eyazici@ku.edu.tr}
\and
Department of Mathematical and Geospatial Sciences, School of Science, \\RMIT University, Melbourne VIC 3000, Australia\\
\email{asha.rao@rmit.edu.au}
}

\maketitle              

\begin{abstract}
We study investigative subgraph extraction for Anti-Money Laundering (AML) transaction graphs using the Shortest Paths Network Search Algorithm (SPNSA), which aggregates shortest paths among a small input feed of entities. Since SPNSA is highly sensitive to the feed choice, we evaluate several graph-only feed selection strategies and propose a motif-based coherent feed selection method. Experiments on the IBM-AML benchmark (\textsc{HI-Small}, \textsc{LI-Small}, \textsc{HI-Medium}) and the \textsc{Elliptic} Bitcoin transaction graph show that the proposed strategy yields markedly higher illicit ratios in the extracted subgraph than random and degree-based heuristics on \textsc{Elliptic} and the IBM-AML HI variants, while keeping the output compact.
\end{abstract}

\section{Introduction}
\label{sec:intro}

Money laundering---the process of hiding the illicit origin of funds through complex sequences of transfers---poses an ongoing and evolving challenge for financial institutions. Modern laundering schemes generate multi-step trails naturally represented as directed transaction graphs, where vertices represent entities (e.g., accounts) and directed edges represent transactions. Identifying these patterns is difficult because illicit activity is 
rare, resulting in highly imbalanced datasets that are difficult to analyze without significant prior information \cite{altman2023realistic,blanusa2024gfp}.

A long-standing barrier in Anti-Money Laundering (AML) research has been the lack of shareable, realistic transaction graphs due to strict privacy and regulatory constraints \cite{lopez2016paysim,pavlidis2020bank}. Recently, Altman et al.\ \cite{altman2023realistic} addressed this by introducing \emph{AMLworld}, an agent-based simulator, and releasing the IBM-AML benchmark datasets. Complementing these synthetic datasets, the \textsc{Elliptic} Bitcoin dataset provides a real-world transaction graph with labeled illicit/licit nodes \cite{Weber2019Elliptic}. Together, these datasets make it feasible to study graph-based investigative methods on both controlled simulation and real transaction datasets.

Much of recent literature using the IBM-AML benchmark datasets focuses on optimizing detection models (i.e., classifiers) to handle extreme class imbalance and directed multigraph structures \cite{altman2023realistic,egressy2024provably}. For instance, Egressy et al.\ \cite{egressy2024provably} propose theoretically grounded adaptations that make Graph Neural Networks much more effective on the simulated AML datasets. Blanu\v{s}a et al.\ \cite{blanusa2024gfp} introduce the \emph{Graph Feature Preprocessor} to mine laundering motifs for edge-level features, significantly reducing runtime while maintaining accuracy. Other studies explore privacy-preserving prediction via homomorphic encryption \cite{effendi2024privacy} and streaming learning for evolving fraud patterns \cite{deprez2025continual}.

In contrast, our work focuses on a task different from supervised detection. We study investigative subgraph extraction, specifically the strategic selection of the algorithm feed. Rather than proposing a new classifier, we aim to produce a compact investigative subgraph with a high proportion of illicit nodes (i.e., high illicit ratio), an essential tool for human analysts who cannot feasibly inspect a global transaction graph. We evaluate this objective on different variants of IBM-AML and \textsc{Elliptic} datasets.

To this end, we build upon the Shortest Paths Network Search Algorithm (SPNSA) proposed by Magalingam et al. \cite{magalingam2015spnsa}. SPNSA takes an input ``feed'' $F \subseteq V$ (a small set of entities of interest) and constructs a subgraph by aggregating feed-guided shortest paths. For each ego node $s \in F$, SPNSA explores the $r$-hop neighborhood, identifies structurally significant intermediate nodes via centrality measures, and extracts the paths connecting $s$ to these nodes. This approach yields subgraphs significantly more interpretable than naive $k$-hop expansions that often grow rapidly and become difficult to manage in dense networks \cite{magalingam2015spnsa}.

However, SPNSA has a critical limitation: its output quality is highly sensitive to the choice of $F$. If the feed nodes are poorly chosen or scattered, the algorithm may route paths through high-connectivity nodes (the financial equivalent of noise) that are not semantically relevant, reducing the illicit ratio of the output.

In addition, the original SPNSA study evaluates the method on the Enron e-mail communication graph from the early 2000s \cite{klimt2004enron,magalingam2015spnsa}, which is both domain-specific and temporally distant from modern transaction ecosystems. The study \cite{magalingam2015spnsa} chooses algorithm feeds based on organizational roles in Enron (e.g., selecting top managers) providing an intuitive way to instantiate $F$.
In general transaction datasets, such role-based intuition is typically unavailable: we rarely know beforehand which nodes correspond to managers, intermediaries, or suspected launderers; feeds chosen naively could lead SPNSA to extract mostly benign, uninformative subgraphs.
This motivates our central research question: \emph{how do we select a feed strategically, using only graph-available signals, such that SPNSA yields a small investigative subgraph enriched with illicit activity?}

We study SPNSA in modern transaction graphs and focus on the problem of selecting the algorithm feed $F$ using only graph-available signals.
We evaluate several practical feed selection strategies and systematically analyze how the choice of $F$ affects both the size and illicit enrichment of the extracted SPNSA subgraph across IBM-AML and \textsc{Elliptic} datasets.
Based on these findings, we propose a new motif-based feed selection algorithm that outperforms the other tested strategies and achieves substantial gains on IBM-AML HI variants as well as an approximately $14\times$ higher illicit ratio than random feed selection on \textsc{Elliptic}.

The remainder of the paper is organized as follows. Section~\ref{sec:methodology} describes SPNSA and the feed selection strategies, including our motif-based method. Section~\ref{sec:experiments} presents the datasets, preprocessing, and evaluation protocol. Section~\ref{sec:illicit_ratio} reports the experimental results, and Section~\ref{sec:discussion} discusses implications and limitations. Finally, Section~\ref{sec:conclusion} concludes the paper and outlines future work.

\footnotesize \textbf{Code availability.} To support reproducibility, we release the implementation at \url{https://github.com/koshimbetovv/spnsa-feed-selection}.

\section{Methodology}
\label{sec:methodology}

\subsection{Shortest Paths Network Search Algorithm (SPNSA)}
\label{sec:spnsa}


SPNSA is a feed-guided subgraph extraction method designed to produce a compact investigative subgraph when provided with a small, semantically meaningful feed \cite{magalingam2015spnsa}. Given a directed graph $G=(V,E)$ and a feed $F \subseteq V$, SPNSA processes each feed node $s\in F$ as an \emph{ego} node. In our implementation, the ego-network of $s$ is the directed out-ego-network within hop radius $r$, i.e., the induced subgraph on nodes reachable from $s$ by a directed path of length at most $r$ (the \emph{SPNSA radius parameter}). Within this ego-network, SPNSA computes directed betweenness and eigenvector centrality, and selects the highest-scoring nodes as intermediaries. It then adds shortest paths inside the ego-network from $s$ to these intermediary nodes, from $s$ to other feed nodes present in the ego-network, and from those feed nodes to the intermediary nodes. The union of these paths over all feed nodes forms the extracted shortest path network (SPN).

\begin{algorithm}[H]
\caption{SPNSA (feed-guided shortest paths subgraph extraction), adapted from \cite{magalingam2015spnsa}.}
\scriptsize
\label{alg:spnsa}
\begin{algorithmic}[1]
\Require Graph $G=(V,E)$, feed set $F\subseteq V$, hop radius $r$.
\Ensure Extracted subgraph $H=(V_H,E_H)$.
\State $E_H \gets \emptyset$, $V_H \gets \emptyset$
\ForAll{$s \in F$}
    \State $G_s \gets$ ego-network of $s$ in $G$ within $r$ hops
    \State Select intermediary nodes $M_s \gets$ central nodes in $G_s$ (max betweenness and max eigenvector)
    \State $T_s \gets (M_s \cup (F \cap V(G_s)))$
    \ForAll{$t \in T_s$}
        \State $P \gets$ a shortest path from $s$ to $t$ in $G_s$ (if exists)
        \State $E_H \gets E_H \cup E(P)$; \; $V_H \gets V_H \cup V(P)$
    \EndFor
\EndFor
\State \Return $H = G[V_H]$ with edges $E_H$
\end{algorithmic}
\end{algorithm}

\subsection{Feed construction strategies}
\label{sec:feed_strategies}

We compare multiple strategies for constructing a feed $F$ of size $k$.
Strategies use only graph topology and simple degree statistics (no illicit labels), except S0 which is a label-informed upper bound used only for reference.

\noindent\textbf{S0: High illicit degree (label-informed upper bound).}
Rank nodes by their number of connections to known illicit nodes and take top-$k$.

\noindent\textbf{S1: Uniform random.}
Sample $k$ nodes uniformly at random from the graph.

\noindent\textbf{S2: Largest degree difference.}
Rank nodes by $|\!\deg^{out}(v)-\deg^{in}(v)|$ and take the top-$k$ nodes.

\noindent\textbf{S3: Mixed collect/distribute.}
Select two groups of nodes:
(i) nodes with $\deg^{in}(v)>0$ and $\deg^{out}(v)=0$ (only receivers), taking the top $k/2$ nodes by $\deg^{in}$; and
(ii) nodes with $\deg^{out}(v)>0$ and $\deg^{in}(v)=0$ (only senders), taking the top $k/2$ by $\deg^{out}$.
The final feed set is the union of these two groups.

\begin{figure}[!htbp]
    \centering
    \includegraphics[width=0.9\linewidth]{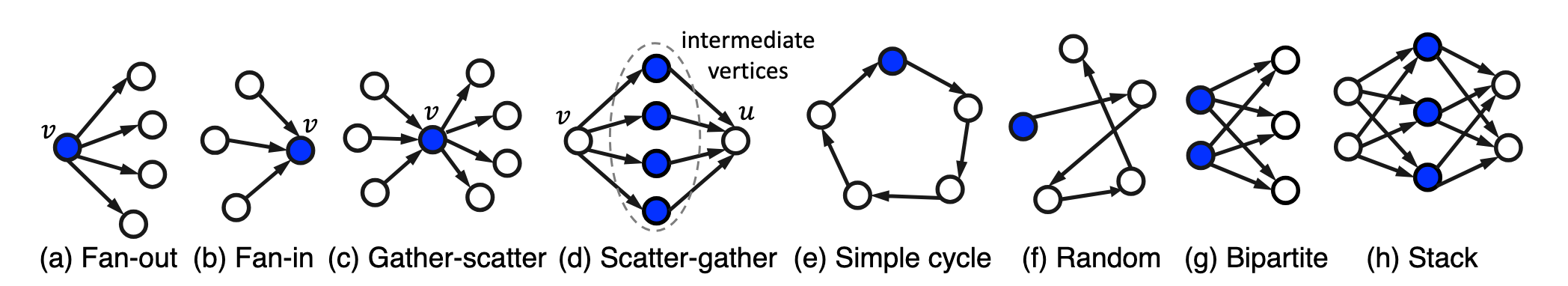}
    \caption{\footnotesize Local transaction motifs used to model common Money Laundering typologies. Blue nodes indicate key intermediaries in each pattern. Reproduced from \cite{altman2023realistic}.}
    \label{fig:aml_motifs}
\end{figure}

\noindent\textbf{S4: Motif-based coherent feed selection.}
\label{sec:motif_feed}
SPNSA is feed-sensitive: if feed nodes are dispersed, shortest paths can pass through high-connectivity regions, producing large subnetworks with low illicit ratio.
Moreover, to reveal illicit structure, the feed should include at least some illicit nodes allowing SPNSA to expand around their neighborhoods and connect them to other suspicious nodes.
To address both issues, we propose a motif-based feed selection strategy (S4) targeting nodes likely to participate in laundering patterns visible in graph structure alone, such as short cycles (round-tripping), fan-in/fan-out (aggregation/distribution), and gather--scatter or scatter--gather patterns involving intermediate ``relay'' vertices linking multiple incoming and outgoing flows \cite{altman2023realistic,blanusa2024gfp}.
Figure~\ref{fig:aml_motifs} illustrates standard transaction motifs commonly used to describe such typologies.

These local patterns suggest that nodes participating in such motifs are more informative feeds than purely random nodes. Accordingly, S4 assigns each node a \emph{motif-based suspicion score} $s(v)$ (see Eqn. \ref{eq:motif_score}) using only the local directed neighborhood structure.
Nodes with high $s(v)$ are treated as promising feeds.
Because computing motif statistics for every node can be expensive at scale, we restrict attention to a candidate pool and compute $s(v)$ only within that pool.

\textbf{Candidate pool (scalability).}
We construct a candidate pool $\mathcal{C}$ by selecting the top $C$ nodes from the full graph according to a simple scoring rule
$$\deg^{in}(v)\cdot\deg^{out}(v),$$ which prioritizes nodes with both incoming and outgoing activity. We then compute the full motif-based score $s(v)$ only for $v\in\mathcal{C}$.

\textbf{Motif scoring.}
Let $G=(V,E)$ be the directed transaction graph, where $V$ is the set of accounts and
$E \subseteq V\times V$ is the set of directed transaction edges.
Let $\Gamma^{in}(v)=\{u:(u\to v)\in E\}$ and $\Gamma^{out}(v)=\{w:(v\to w)\in E\}$.
We compute the following motif counts:
\begin{align*}
\textsf{recip}(v)  &:= |\Gamma^{in}(v)\cap \Gamma^{out}(v)| && \text{(reciprocal transfers)}, \\
\textsf{relay}(v)  &:= |\Gamma^{in}(v)|\cdot|\Gamma^{out}(v)| && \text{(pass-through/intermediate potential)}, \\
\textsf{cycle3}(v) &:= \sum_{w\in \Gamma^{out}(v)} |\Gamma^{out}(w)\cap \Gamma^{in}(v)|
&& \text{(directed 3-cycles $u\!\to\!v\!\to\!w\!\to\!u$)} .
\end{align*}
To capture star-like aggregation/distribution while avoiding generic high-degree hubs, we use a thresholded star indicator with star ratio parameter $\tau$, whose dataset-specific values are reported in Table~\ref{tab:S4_hyperparams}:
\begin{equation}
    \label{eq:fan_in_out}
    \begin{aligned}
    \textsf{fanin}(v) &:= |\Gamma^{in}(v)|\cdot \mathbb{I}\bigg[ \frac{|\Gamma^{in}(v)|}{|\Gamma^{out}(v)|+1} \ge \tau \bigg],\\
    \textsf{fanout}(v)&:= |\Gamma^{out}(v)|\cdot \mathbb{I}\bigg[ \frac{|\Gamma^{out}(v)|}{|\Gamma^{in}(v)|+1} \ge \tau \bigg],
    \end{aligned}
\end{equation}
where $\mathbb{I}[\cdot]$ is the indicator function (equal to $1$ if the condition holds, and $0$ otherwise).
Intuitively, $\textsf{fanin}(v)$ (resp.\ $\textsf{fanout}(v)$) equals the in-degree (resp.\ out-degree) of $v$ only when $v$ is sufficiently imbalanced toward incoming (resp.\ outgoing) edges by at least a factor $\tau$; otherwise it is set to $0$.
This helps avoid flagging generic hubs, since nodes with large in- and out-degree typically have ratios close to $1$ and therefore fail the $\tau>1$ threshold, yielding zero score unless their connectivity is strongly one-sided (i.e., star-like).

To further distinguish suspicious star-like behavior from generic hubs, we also count how many neighbors of $v$ are \emph{leaf-like}, i.e., have very small total degree. Let $d_{leaf}$ denote the maximum total degree for a node to be considered leaf-like. We define
\begin{equation}
\label{eq:leaf_star}
\textsf{leaf\_star}(v) := |\{u \in \Gamma^{in}(v) : \deg(u) \le d_{leaf}\}| + |\{w \in \Gamma^{out}(v) : \deg(w) \le d_{leaf}\}|,
\end{equation}
where $\deg(x)=|\Gamma^{in}(x)|+|\Gamma^{out}(x)|$ denotes total degree. Intuitively, $\textsf{leaf\_star}(v)$ is large when $v$ is attached to many low-degree accounts, which is consistent with star-shaped aggregation/distribution patterns built around peripheral accounts.

We then combine these signals with log-compression and an explicit degree penalty to get $s(v)$:
\begin{equation}
\label{eq:motif_score}
\begin{aligned}
s(v) \;=\;& \log\!\big(1+\textsf{cycle3}(v)\big)
+\alpha\,\log\!\big(1+\textsf{recip}(v)\big)
+\beta\,\log\!\big(1+\textsf{relay}(v)\big) \\
&+\texttt{s\_in}\log\!\big(1+\textsf{fanin}(v)\big)
+\texttt{s\_out}\log\!\big(1+\textsf{fanout}(v)\big)\\
&+\texttt{s\_leaf}\log\!\big(1+\textsf{leaf\_star}(v)\big) -\gamma\,\log\!\big(1+\deg(v)\big).
\end{aligned}
\end{equation}
Here, $\alpha$, $\beta$, $\gamma$, $\texttt{s\_in}$, $\texttt{s\_out}$, and $\texttt{s\_leaf}$ are nonnegative weighting parameters controlling the contribution of each component(Table~\ref{tab:S4_hyperparams}).
The log terms prevent a small number of extreme counts from dominating, while the $-\log(1+\deg(v))$ term discourages selecting purely generic hubs that would tend to create ``corridor'' shortest paths.

\textbf{Coherence constraint (multi-center selection).}
Let $\pi$ be candidates sorted by decreasing $s(v)$.
Rather than taking the top-$k$ nodes globally, we enforce feed coherence by selecting from multiple suspicious regions:
we choose the top $m$ nodes in $\pi$ as centers, form the undirected projection $U$ of the graph, and for each center $c$ select high-scoring nodes within the ball
$B(c)=\{v:\mathrm{dist}_U(c,v)\le d_{\max}\}$ until the total feed size reaches $k$. Here, $m$, $k$ and $d_{\max}$ are algorithm parameters where $k$ denotes the desired \emph{feed size}, and $d_{\max}$ controls \emph{locality/coherence}.

Using multiple centers reduces the risk of collapsing onto a single region and increases the chance of covering several laundering groups, while local selection increases ego-network overlap and empirically improves illicit enrichment of the SPNSA output. The complete procedure is provided in Algorithm~\ref{alg:motif-feed}.

\begin{algorithm}[h]
\scriptsize
\caption{Motif-based Coherent Feed Selection (S4)}
\label{alg:motif-feed}
\begin{algorithmic}[1]
\Require Directed graph $G=(V,E)$, feed size $k$, candidate pool size $C$, coherence parameters (\texttt{centers} $m$, distance cutoff $d_{\max}$), motif parameters ($\alpha$, $\beta$, $\gamma$, $\tau$, \texttt{s\_in}, \texttt{s\_out}, \texttt{s\_leaf}, $d_{leaf}$)
\Ensure Feed set $F \subseteq V$ with $|F|=k$

\State Compute degrees $\deg^{in}(v), \deg^{out}(v)$ for all $v \in V(G)$
\State Form candidate pool $\mathcal{C}$ by selecting the top-$C$ nodes by a score $\deg^{in}(v)\cdot\deg^{out}(v)$ from $G$

\For{each $v \in \mathcal{C}$}
    \State $In(v) \gets \{u : (u \to v) \in E(G)\}$,\quad $Out(v) \gets \{w : (v \to w) \in E(G)\}$
    \State $\textsf{recip}(v) \gets |In(v) \cap Out(v)|$ \Comment{reciprocal pairs}
    \State $\textsf{relay}(v) \gets |In(v)|\cdot|Out(v)|$ \Comment{pass-through proxy}
    \State $\textsf{cycle3}(v) \gets \sum_{w \in Out(v)} |Out(w) \cap In(v)|$ \Comment{3-cycles $u\!\to\!v\!\to\!w\!\to\!u$}
    \State $\textsf{fanin}(v) \gets |In(v)| \cdot \mathbb{I}\big[|In(v)| \ge \texttt{$\tau$}\cdot(|Out(v)|+1)\big]$
    \State $\textsf{fanout}(v) \gets |Out(v)| \cdot \mathbb{I}\big[|Out(v)| \ge \texttt{$\tau$}\cdot(|In(v)|+1)\big]$
    \State $\deg(v) \gets |In(v)| + |Out(v)|$
    \State $s(v) \gets \log(1+\textsf{cycle3}(v)) + \alpha\log(1+\textsf{recip}(v)) + \beta\log(1+\textsf{relay}(v)) +\texttt{s\_in}\log(1+\textsf{fanin}(v))$
    \State \hspace{2.55cm} $+~\texttt{s\_out}\log(1+\textsf{fanout}(v)) +\texttt{s\_leaf}\log(1+\textsf{leaf\_star}(v)) - \gamma\log(1+\deg(v))$
\EndFor

\State Let $\pi$ be nodes in $\mathcal{C}$ sorted by decreasing $s(v)$
\State Let $U \gets$ undirected projection of $G$
\State Initialize $F \gets \emptyset$
\State Choose centers $A \gets$ first $m$ nodes of $\pi$ (top-scoring centers)

\For{each $c \in A$}
    \State Compute ball $B(c) \gets \{v \in V(U): \mathrm{dist}_U(c,v) \le d_{\max}\}$
    \For{each $v$ in $\pi$}
        \If{$v \in B(c)$ \textbf{and} $v \notin F$}
            \State $F \gets F \cup \{v\}$
            \If{$|F| = k$} \Return $F$ \EndIf
        \EndIf
    \EndFor
\EndFor

\Comment{Fallback: fill remaining slots with top-ranked nodes}
\For{each $v$ in $\pi$}
    \If{$v \notin F$}
        \State $F \gets F \cup \{v\}$
        \If{$|F| = k$} \Return $F$ \EndIf
    \EndIf
\EndFor

\State \Return $F$
\end{algorithmic}
\end{algorithm}

\section{Experiments and Results}
\label{sec:experiments}

\subsection{Datasets}
\label{sec:data}

We evaluate our methods on two transaction datasets: the IBM Anti-Money Laundering benchmark (IBM-AML) \cite{altman2023realistic} and the \textsc{Elliptic} Bitcoin transaction dataset \cite{Weber2019Elliptic}.

\textbf{IBM-AML synthetic benchmark.}
The IBM-AML benchmark provides large synthetic AML transaction graphs generated by an agent-based simulator \cite{altman2023realistic}.
Accounts are modeled as nodes and transactions as directed, timestamped edges.
The benchmark is released at multiple scales (Small/Medium/Large) and activity levels (Higher Illicit ratio, HI / Lower Illicit ratio, LI). 
In our experiments, we use three IBM-AML variants: \textsc{HI-Small}, \textsc{LI-Small}, and \textsc{HI-Medium}.

\textbf{\textsc{Elliptic} Bitcoin transaction dataset.} The \textsc{Elliptic} dataset provides a directed transaction graph derived from Bitcoin transactions, together with illicit/licit labels for a subset of nodes \cite{Weber2019Elliptic}.
We use this dataset as a real-world complement to the synthetic IBM-AML benchmark.

\subsection{Preprocessing}
\label{sec:preprocessing}

The IBM-AML dataset provides illicit labels at the \emph{transaction (edge)} level \cite{altman2023realistic}.
To evaluate illicit enrichment in extracted subgraphs, we derive node labels by marking an account as illicit if it appears as an endpoint of at least one laundering transaction.

In our IBM-AML variants, we observed that $99.6\%$ of illicit-labeled nodes lie in the largest weakly connected component. Therefore, we run all IBM-AML experiments on the largest component to exclude small disconnected components that are not relevant to our evaluation.

In both datasets, we convert the transaction graph to a simple directed graph by keeping a single directed edge $(u,v)$ if at least one transaction from $u$ to $v$ exists. Summary statistics of the preprocessed graphs are reported in Table~\ref{tab:graph_stats_used}.
\vspace{-10pt}
\begin{table}[!htbp]
\centering
\small
\setlength{\tabcolsep}{4pt}
\renewcommand{\arraystretch}{1.1}
\caption{\footnotesize Dataset statistics used in our experiments. $G$ denotes the (preprocessed) transaction graph; $V(G)$ and $E(G)$ are its vertex and edge sets; $I(G)$ is the set of illicit nodes; and $IR(G)=|I(G)|/|V(G)|$ is the illicit ratio.}\vspace{2mm}
\label{tab:graph_stats_used}
{%
\begin{tabular}{lrrrr}
\toprule
\textbf{Dataset} & $\lvert V(G)\rvert$ & $\lvert E(G)\rvert$ & $\lvert I(G)\rvert$ & $IR(G)$ \\
\midrule
\textsc{Elliptic}  & 203769 & 234355 & 4545 & $2.23\%$ \\
\textsc{HI-Small}  & 372091 & 874017 & 6342 & $1.70\%$ \\
\textsc{LI-Small}  & 504391 & 1184998 & 5285 & $1.04\%$ \\
\textsc{HI-Medium} & 1611244 & 4015935 & 41774 & $2.59\%$ \\
\bottomrule
\end{tabular}%
}
\end{table}
\vspace{-25pt}

\subsection{Evaluation protocol}
\label{sec:metrics}

For a feed $F$ and SPNSA radius $r$, let $H(F)$ denote the extracted SPN.
Let $V(H)$ and $E(H)$ denote its vertex and edge sets, and let $I(H)\subseteq V(H)$ denote the set of illicit nodes in $H$.
We report the size of the extracted subgraph as $|V(H)|$ and $|E(H)|$, and we evaluate illicit enrichment using the ratio
\[
IR(H)=\frac{|I(H)|}{|V(H)|}=\frac{|\{v\in V(H):y_v=\text{illicit}\}|}{|V(H)|},
\]
where $y_v$ is the label of $v$. Labels are used \emph{only} for evaluation, not for feed construction.

For \textsc{Elliptic}, we report two enrichment metrics since it has unlabeled nodes. The main text uses $IR(H)$ treating unlabeled nodes conservatively as licit. In addition, we define a labeled-only variant
$
LIR(H)=\frac{|I(H)|}{|V(H)\cap L|},
$
where $L$ is the set of labeled \textsc{Elliptic} nodes. This shows enrichment using only labeled nodes, without lowering the score because of unlabeled nodes.
\vspace{-10pt}
\subsection{Experimental settings}
\label{sec:setup}

We evaluate strategies S0--S4 from Section~\ref{sec:feed_strategies}.
S0 is a label-informed upper bound and is reported only for reference.
We test two feed sizes, $k\in\{200,500\}$, and two SPNSA radius, $r\in\{1,2\}$, to capture the trade-off between interpretability (smaller $H$) and coverage (more illicit nodes captured).
For S1 (uniform random), we report mean results over 100 random trials.

The hyper-parameters for S4 are listed in Table~\ref{tab:S4_hyperparams}. They were chosen by coarse manual tuning to balance illicit enrichment and subgraph compactness. Most parameters were fixed across datasets; only $\tau$, $s_{\mathrm{in}}$, and the number of centers $m$ were adjusted by dataset family.

\begin{table}[!htbp]
\centering
\scriptsize
\setlength{\tabcolsep}{3pt}
\renewcommand{\arraystretch}{1.0}
\caption{\footnotesize Hyper-parameters used for strategy S4 across datasets. Parameters are defined in Eqs.~\eqref{eq:fan_in_out}--\eqref{eq:motif_score} and in Section~\ref{sec:feed_strategies}. }\vspace{2mm}
\label{tab:S4_hyperparams}
\resizebox{0.9\linewidth}{!}{%
\begin{tabular}{lrrrrrrrrrrrr}
\hline
Dataset
& $C$
& $\alpha$
& $\beta$
& $\gamma$
& $\tau$
& \texttt{s\_in}
& \texttt{s\_out}
& \texttt{s\_leaf}
& $d_{leaf}$
& $d_{\max}$
& $m$/\texttt{centers} \\
\hline
\textsc{Elliptic}   & $300\text{k}$ & $0.2$ & $0.3$ & $0.8$ & $15.0$ & $0.5$ & $0.4$ & $0.6$ & $2$ & $2$ & $7$ \\
\textsc{HI-Medium/Small}  & $300\text{k}$ & $0.2$ & $0.3$ & $0.8$ & $15.0$ & $0.3$ & $0.4$ & $0.6$ & $2$ & $2$ & $5$ \\
\textsc{LI-Small}   & $300\text{k}$ & $0.2$ & $0.3$ & $0.8$ & $27.0$ & $0.2$ & $0.4$ & $0.6$ & $2$ & $2$ & $5$ \\
\hline
\end{tabular}%
}
\vspace{-5mm}
\end{table}

\subsection{Results}
\label{sec:illicit_ratio}

\begin{figure}[!htbp]
\vspace{-5mm}
\centering
\includegraphics[width=1.0\linewidth]{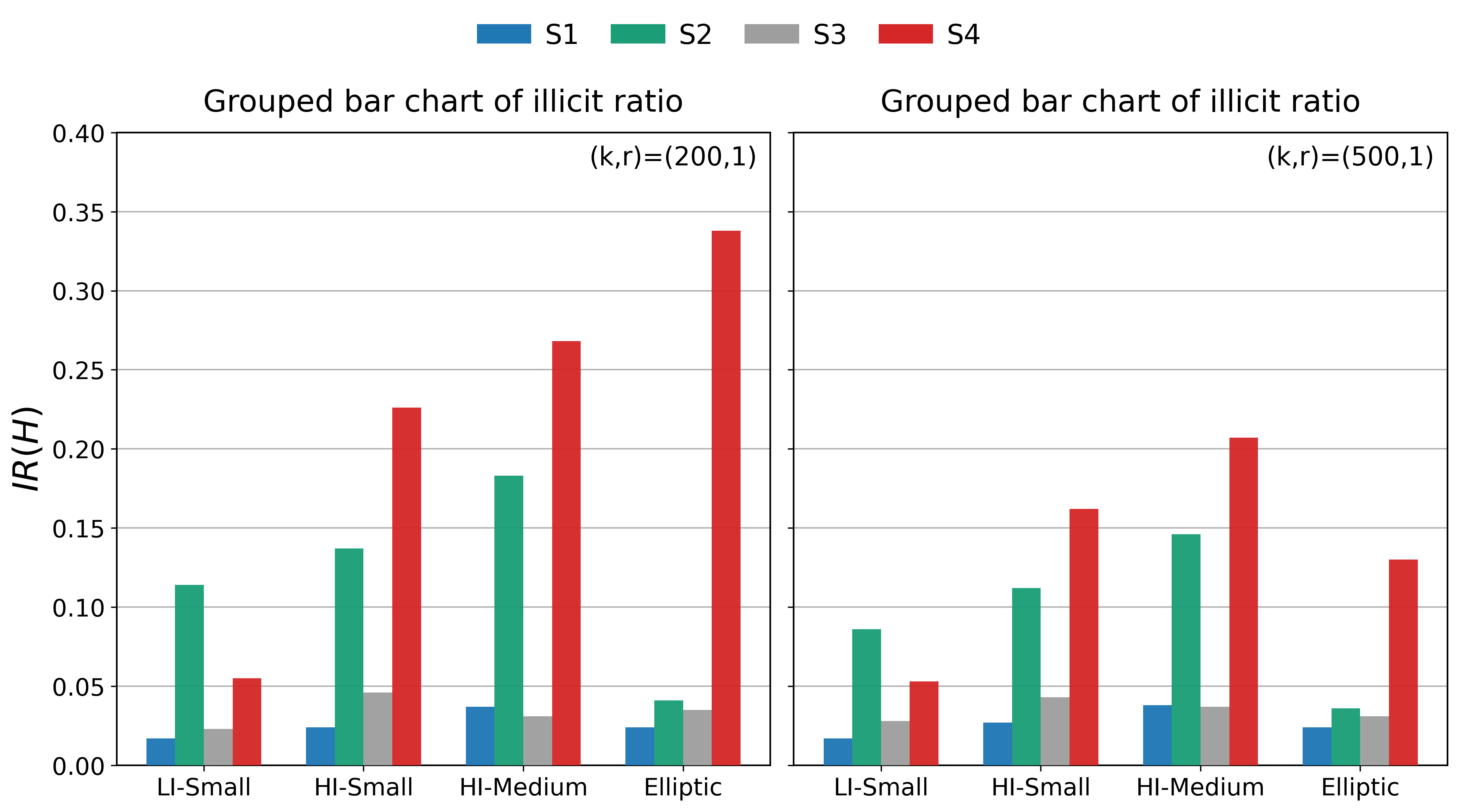}
\caption{\footnotesize Grouped bar charts of illicit ratio $IR(H)$ for feed selection strategies S1--S4 on the four datasets, shown for two SPNSA settings: $(k,r)=(200,1)$ and $(500,1)$, where $k$ is the feed size and $r$ is the ego-network radius. Each group corresponds to a dataset, and bars compare $IR(H)$ across strategies.}
\label{fig:grouped_bar}
\vspace{-10mm}
\end{figure}

\begin{figure}[!htbp]
\centering
\includegraphics[width=1.0\linewidth]{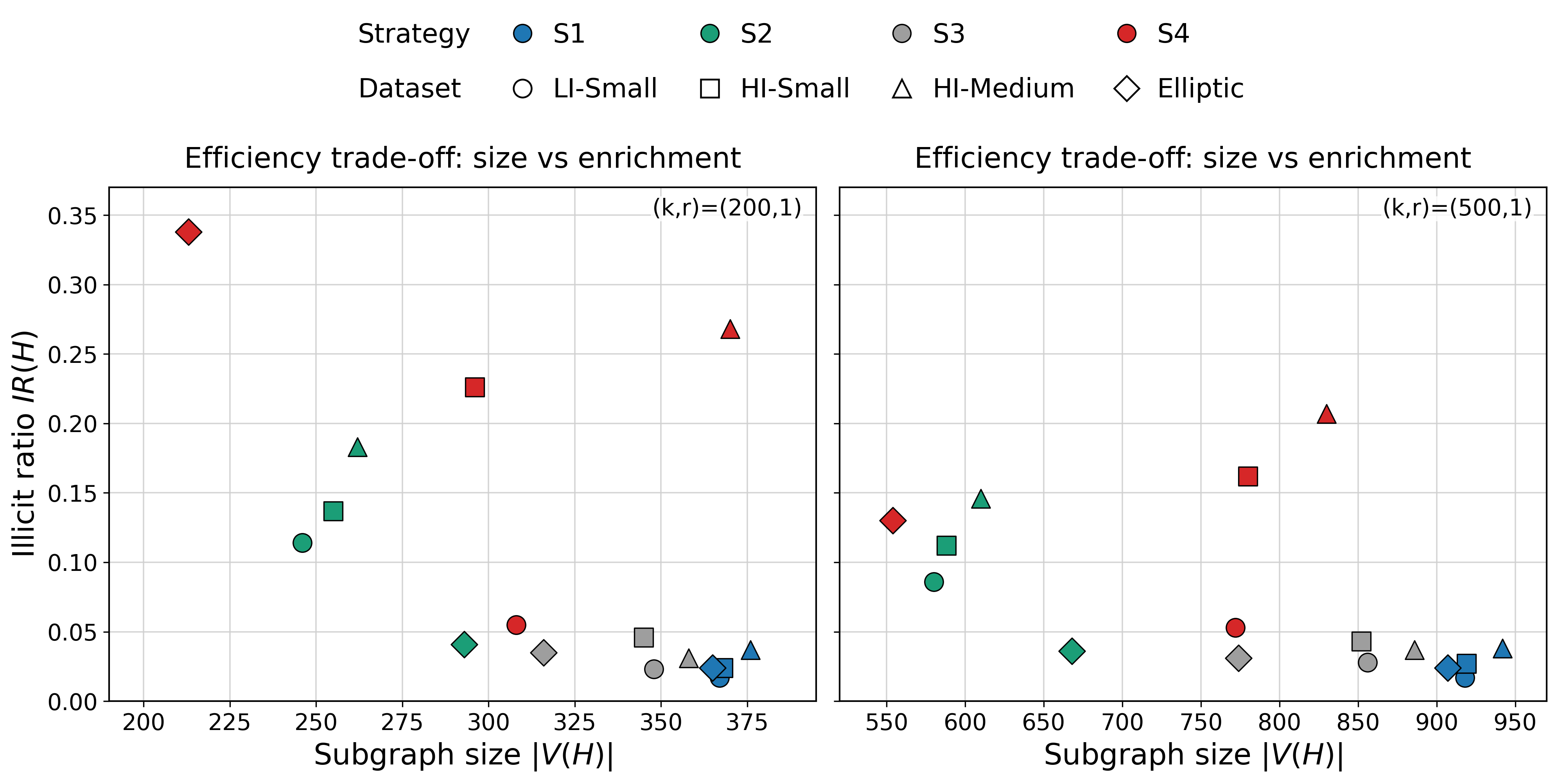}
\caption{\footnotesize Efficiency trade-off between extracted subgraph size $|V(H)|$ and illicit ratio $IR(H)$ for feed selection strategies S1--S4 on four datasets, shown for two SPNSA settings: $(k,r)=(200,1)$ and $(500,1)$, where $k$ is the feed size and $r$ is the ego-network radius. Colors denote strategies and marker shapes denote datasets; points toward the upper-left indicate a better trade-off.}
\label{fig:efficiency_tradeoff}
\vspace{-5mm}
\end{figure}

Table~\ref{tab:feed_strategies_all},\ref{tab:s4_kne_induced_spnsa} in Appendix~\ref{app:results_table} reports all outcomes; we summarize the main patterns below and highlight them with Figures~\ref{fig:grouped_bar} and~\ref{fig:efficiency_tradeoff}.

On the \textsc{Elliptic} Bitcoin transaction graph and the two IBM-AML \textsc{HI} variants, the motif-based coherent feed strategy (S4) achieves the highest illicit ratio for every tested SPNSA configuration.
Concretely, at $(k,r)=(200,1)$ on \textsc{Elliptic}, S4+SPNSA extracts a compact subgraph with only $213$ nodes while concentrating illicit activity to an illicit ratio of $33.8\%$ (Table~\ref{tab:feed_strategies_all}(d)).
This is far above both the dataset base rate ($2.23\%$, Table~\ref{tab:graph_stats_used}) and the uniform-random feed baseline (S1), which remains near $2$--$3\%$ across settings.
A similar picture holds on IBM-AML \textsc{HI-Small} and \textsc{HI-Medium}: S4 raises the illicit ratio into the $\approx 16\%$--$29\%$ range (depending on $(k,r)$) while keeping the extracted subgraph in the few-hundred-node regime (Table~\ref{tab:feed_strategies_all}(b,c)).

On IBM-AML \textsc{LI-Small}, which has the lowest dataset-level illicit ratio ($1.04\%$, Table~\ref{tab:graph_stats_used}), the best feed heuristic is the degree-imbalance strategy (S2), not the motif-based strategy.
For example, at $(k,r)=(200,1)$, S2 achieves an illicit ratio of $11.4\%$, while S4 reaches $5.5\%$ (Table~\ref{tab:feed_strategies_all}(a)).
Thus, even though S4 remains clearly better than random, the strongest \emph{structure-only} indicator in this low-illicit setting appears to be captured by simple in/out-degree asymmetry.

Across datasets and strategies, increasing the feed size from $k=200$ to $k=500$ predictably increases the extracted subgraph size (more vertices and edges) and often decreases the illicit ratio.
A representative example is \textsc{Elliptic} under S4: moving from $(k,r)=(200,1)$ to $(500,1)$ increases the extracted subgraph from $213$ to $554$ nodes, while the illicit ratio drops from $33.8\%$ to $13.0\%$ (Table~\ref{tab:feed_strategies_all}(d)).
Increasing the SPNSA radius from $r=1$ to $r=2$ similarly enlarges the ego-networks and tends to increase the output size; its effect on illicit ratio is dataset-dependent (e.g., it slightly improves S4 on \textsc{HI-Medium}, Table~\ref{tab:feed_strategies_all}(c)).
Overall, these patterns reflect a practical choice: smaller $(k,r)$ typically yields more interpretable, more illicit-concentrated subnetworks, while larger $(k,r)$ increases coverage at the cost of concentration.

Figure~\ref{fig:efficiency_tradeoff} jointly considers extracted subgraph size and illicit ratio.
For \textsc{Elliptic} and the \textsc{HI} variants, S4 tends to occupy a favorable region of this plane: compared to random feeds (S1) and the simpler degree-based heuristics (S2/S3), it attains higher illicit ratios without requiring larger subnetworks.
On \textsc{LI-Small}, the same figure supports the earlier observation that S2 provides the better size--enrichment balance.

\textbf{Note on the source of enrichment.}
To isolate the contribution of feed selection from the subsequent SPNSA aggregation, we also evaluated the subgraph induced by the S4 feed nodes.
Using the \emph{same} S4 feed, the induced subgraph consistently attains an illicit ratio that is equal to or higher than the SPNSA output (Table~\ref{tab:s4_kne_induced_spnsa}, Appendix).
This indicates that the primary enrichment gains attributed to S4+SPNSA arise from the feed construction itself, while SPNSA mainly adds path-based connector nodes and occasionally additional illicit nodes, which can dilute illicit concentration.

We also compared against a simple $k$-hop neighbourhood expansion (kNE) baseline, defined as the union of the $k$-hop ego-neighbourhoods of the feed nodes.
Across all datasets, this expansion yields substantially larger subgraphs yet consistently lower illicit ratios than the induced-feed subgraph and SPNSA (Table~\ref{tab:s4_kne_induced_spnsa}), indicating that naive neighbourhood expansion dilutes enrichment even more strongly.

\vspace{-5pt}
\section{Discussion}
\label{sec:discussion}
\vspace{-5pt}
SPNSA constructs its output by aggregating shortest paths found inside ego-networks around the feed nodes.
If the feed is chosen without regard to laundering-relevant structure, these ego-networks are more likely to cover unrelated areas of the graph, and the union of the resulting paths can include many benign vertices, reducing the illicit ratio.
The motif-based coherent strategy (S4) targets the opposite behavior: it prioritizes nodes whose local directed neighborhoods are similar to common laundering typologies, and then enforces locality through multi-center selection.
Empirically, this combination shifts SPNSA toward extracting subnetworks that are both \emph{compact} and \emph{enriched}, which is exactly the desired outcome for human-centered investigation (Figures~\ref{fig:grouped_bar} and~\ref{fig:efficiency_tradeoff}).

The \textsc{LI-Small} results show that not every dataset rewards the same structural signal.
In that low-illicit dataset setting, the degree-imbalance heuristic (S2) provides the strongest enrichment (Table~\ref{tab:feed_strategies_all}(a)).
One plausible interpretation is that, when illicit activity is sparse, the most informative feed nodes may be those that act as mostly senders or mostly receivers, whereas the specific motif score used by S4 (with the current hyper-parameters) is not as well matched to the dominant illicit structures.
This suggests a practical takeaway: feed selection should be treated as a dataset-dependent design choice rather than a fixed preprocessing step.

Across all strategies, larger feeds and larger radius increase the output size and can reduce illicit concentration.
For investigative use, the smaller settings (e.g., $k=200$ and $r=1$) are attractive starting points.
When the goal shifts toward broader coverage, increasing $k$ or $r$ is natural, but the results indicate that one should expect a reduction in illicit ratio.

\textbf{Limitations.}
First, on \textsc{Elliptic}, the main illicit-ratio metric treats unlabeled nodes as licit for evaluation, which can underestimate the true illicit ratio; to mitigate this, we also report a labeled-only variant.
Second, we simplify transaction multigraphs into simple directed graphs and ignore timestamps during graph construction, removing transaction frequency and temporal information that may be relevant to laundering behavior.

\vspace{-5pt}

\section{Conclusion and Future Work}
\label{sec:conclusion}
\vspace{-5pt}
This work studied investigative subgraph extraction in transaction networks through SPNSA, with a focus on a practically central question: how can we select the feed using only graph-available signals, so that SPNSA returns a small investigative subgraph with a high share of illicit activity?
We show that this choice strongly changes what SPNSA returns.
Across the real \textsc{Elliptic} network and the simulated IBM-AML higher-illicit benchmarks (\textsc{HI-Small/Medium}), our motif-guided coherent feed selection (S4) consistently produces subnetworks with much higher illicit ratios than random feeds and common degree-based heuristics, while keeping the extracted subgraphs small enough to support human inspection.
At the same time, the IBM-AML lower-illicit setting (\textsc{LI-Small}) results show that the best graph-only strategy can depend on the dataset regime, and that a simple degree-difference rule can outperform S4 in the low-illicit setting.

Future work includes incorporating temporal and multi-transaction information into both motif scoring and SPNSA aggregation, developing adaptive (data-driven) parameter selection for motif-based feed construction, and validating the findings on additional real transaction graphs as they become available.

\appendix

\section{Appendix}
\label{app:results_table}
\newcommand{\spnsaPanelHeader}[1]{%
  \vspace{2mm}
  \textbf{#1}\par\vspace{1mm}
}
\vspace{-15pt}
\begin{table}[!htbp] 
\centering
\scriptsize
\setlength{\tabcolsep}{2.5pt}
\renewcommand{\arraystretch}{1.05}
\vspace{-7mm}
\caption{\footnotesize SPNSA results under different feed selection strategies for four datasets:
IBM-AML \textsc{LI-Small}, \textsc{HI-Small}, \textsc{HI-Medium}, and \textsc{Elliptic}.
$H$ denotes the extracted SPN; $V(H)$ and $E(H)$ are its vertex and edge sets, respectively;
$I(H)$ is the set of illicit nodes in $H$; and $IR(H)=|I(H)|/|V(H)|$ is the illicit ratio.
Here, $k$ denotes the feed size and $r$ denotes the radius of ego-network used in SPNSA.
For S1, values are averaged over 100 random trials.}
\label{tab:feed_strategies_all}

\begin{minipage}[t]{0.43\textwidth}
\centering
\spnsaPanelHeader{(a) IBM-AML (\textsc{LI-Small})}
\resizebox{\linewidth}{!}{%
\begin{tabular}{crrrrrr}
\hline
 & $k$ & $r$ & $|V|$ & $|E|$ & $|I|$ & $IR$ \\
\hline
S0 & 200 & 1 & 449 & 397 & 261 & 0.581 \\
S1 & 200 & 1 & 367 & 168 & 6.4  & 0.017 \\
S2 & 200 & 1 & 246 & 134 & 28 & \textbf{0.114} \\
S3 & 200 & 1 & 348 & 148 & 8 & 0.023 \\
S4 & 200 & 1 & 308 & 540 & 17  & 0.055 \\
\hline
S0 & 200 & 2 & 612 & 600 & 313 & 0.511 \\
S1 & 200 & 2 & 454 & 257  & 7.8  & 0.017 \\
S2 & 200 & 2 & 309 & 235 & 30 & \textbf{0.097} \\
S3 & 200 & 2 & 476 & 276 & 11 & 0.023 \\
S4 & 200 & 2 & 317 & 554 & 15  & 0.047 \\
\hline
S0 & 500 & 1 & 1027 & 996 & 594 & 0.578 \\
S1 & 500 & 1 & 918 & 420 & 16.0 & 0.017 \\
S2 & 500 & 1 & 580 & 329 & 50 & \textbf{0.086} \\
S3 & 500 & 1 & 856 & 367 & 24 & 0.028 \\
S4 & 500 & 1 & 772 & 1181 & 41 & 0.053 \\
\hline
\end{tabular}%
}
\end{minipage}
\hspace{10mm}
\begin{minipage}[t]{0.41\textwidth}
\centering
\spnsaPanelHeader{(b) IBM-AML (\textsc{HI-Small})}
\resizebox{\linewidth}{!}{%
\begin{tabular}{crrrrrr}
\hline
 & $k$ & $r$ & $|V|$ & $|E|$ & $|I|$ & $IR$ \\
\hline
S0 & 200 & 1 & 423 & 394 & 307 & 0.726 \\
S1 & 200 & 1 & 368 & 168 & 8.83  & 0.024 \\
S2 & 200 & 1 & 255 & 134 & 35 & 0.137 \\
S3 & 200 & 1 & 345 & 147 & 16 & 0.046 \\
S4 & 200 & 1 & 296 & 456 & 67  & \textbf{0.226} \\
\hline
S0 & 200 & 2 & 660 & 778 & 448 & 0.679 \\
S1 & 200 & 2 & 448 & 251  & 13.7  & 0.031 \\
S2 & 200 & 2 & 326 & 249 & 41 & 0.126 \\
S3 & 200 & 2 & 467 & 270 & 21 & 0.045 \\
S4 & 200 & 2 & 313 & 494 & 69  & \textbf{0.220} \\
\hline
S0 & 500 & 1 & 944 & 960 & 678 & 0.718 \\
S1 & 500 & 1 & 919 & 422 & 24.41 & 0.027 \\
S2 & 500 & 1 & 588 & 320 & 66 & 0.112 \\
S3 & 500 & 1 & 852 & 368 & 37 & 0.043 \\
S4 & 500 & 1 & 780 & 987 & 126 & \textbf{0.162} \\
\hline
\end{tabular}%
}
\end{minipage}

\end{table}

\vspace{-35pt}

\begin{table}[!htbp] 
\centering
\scriptsize
\setlength{\tabcolsep}{2.5pt}
\renewcommand{\arraystretch}{1.05}

\begin{minipage}[t]{0.43\textwidth}
\centering
\spnsaPanelHeader{(c) IBM-AML (\textsc{HI-Medium})}
\resizebox{\linewidth}{!}{%
\begin{tabular}{crrrrrr}
\hline
 & $k$ & $r$ & $|V|$ & $|E|$ & $|I|$ & $IR$ \\
\hline
S0 & 200 & 1 & 431 & 564 & 338  & 0.784 \\
S1 & 200 & 1 & 376 & 176 & 14.07  & 0.037 \\
S2 & 200 & 1 & 262 & 200 & 48 & 0.183 \\
S3 & 200 & 1 & 358 & 158 & 11 & 0.031 \\
S4 & 200 & 1 & 370 & 907 & 99  & \textbf{0.268} \\
\hline
S0 & 200 & 2 & 756 & 1100 & 590  & 0.780 \\
S1 & 200 & 2 & 462 & 265  & 21.8  & 0.047 \\
S2 & 200 & 2 & 319 & 299 & 63 & 0.198 \\
S3 & 200 & 2 & 459 & 259 & 28 & 0.061 \\
S4 & 200 & 2 & 377 & 973 & 108  & \textbf{0.287} \\
\hline
S0 & 500 & 1 & 1040 & 1369 & 758 & 0.729 \\
S1 & 500 & 1 & 942 & 442 & 35.5 & 0.038 \\
S2 & 500 & 1 & 610 & 450 & 89 & 0.146 \\
S3 & 500 & 1 & 886 & 386 & 33 & 0.037 \\
S4 & 500 & 1 & 830 & 1623 & 172 & \textbf{0.207} \\
\hline
\end{tabular}%
}
\end{minipage}
\hspace{10mm}
\begin{minipage}[t]{0.41\textwidth}
\centering
\spnsaPanelHeader{(d) Elliptic: At evaluation, IR treats unlabeled nodes as licit, whereas LIR excludes unlabeled nodes from the denominator.}
\resizebox{\linewidth}{!}{%
\begin{tabular}{crrrrrrr}

\hline
 & $k$ & $r$ & $|V|$ & $|E|$ & $|I|$ & $IR$ & $LIR$ \\
\hline
S0 & 200 & 1 & 334 & 191 & 322 & 0.964 & 0.976\\
S1 & 200 & 1 & 365 & 166 & 8.57  & 0.024 & 0.086 \\
S2 & 200 & 1 & 293 & 140 & 12 & 0.041 & 0.057 \\
S3 & 200 & 1 & 316 & 157 & 11 & 0.035 & 0.050 \\
S4 & 200 & 1 & 213 & 209 & 72  & \textbf{0.338} & \textbf{0.800} \\
\hline
S0 & 200 & 2 & 416 & 309 & 386 & 0.928 & 0.952 \\
S1 & 200 & 2 & 498 & 301  & 10.3  & 0.021 & 0.073 \\
S2 & 200 & 2 & 351 & 232 & 12 & 0.034 & 0.049 \\
S3 & 200 & 2 & 411 & 296 & 12 & 0.029 & 0.042 \\
S4 & 200 & 2 & 218 & 214 & 71  & \textbf{0.326} & \textbf{0.750} \\
\hline
S0 & 500 & 1 & 625 & 498 & 592 & 0.947 & 0.964 \\
S1 & 500 & 1 & 907 & 414 & 21.73 & 0.024 & 0.082 \\
S2 & 500 & 1 & 668 & 372 & 24 & 0.036 & 0.052 \\
S3 & 500 & 1 & 774 & 454 & 24 & 0.031 & 0.049 \\
S4 & 500 & 1 & 554 & 691 & 72 & \textbf{0.130} & \textbf{0.355} \\
\hline
\end{tabular}%
}
\end{minipage}

\end{table}

\begin{table*}[t]
\centering
\caption{\footnotesize Comparison of S4-based subgraphs obtained by k-hop neighborhood expansion (kNE), induced-subgraph extraction, and SPNSA across datasets (kNE k=1; SPNSA radius r=1; feed size 200).}
\label{tab:s4_kne_induced_spnsa}
\resizebox{0.9\textwidth}{!}{%
\begin{tabular}{llrrrrr}
\toprule
\textbf{Dataset} & \textbf{Strategy} & \textbf{$|V|$} & \textbf{$|E|$} & \textbf{$|I|$} & \textbf{IR} & \textbf{Components} \\
\midrule
\multirow{3}{*}{Elliptic}
& S4 kNE subgraph     & 226   & 222    & 72.0   & 0.319 & 5  \\
& S4 induced subgraph & 200   & 195    & 71.0   & \textbf{0.355} & 6  \\
& S4 SPNSA subgraph   & 213   & 209    & 72.0   & 0.338 & 5  \\
\midrule
\multirow{3}{*}{IBM-AML (LI-Small)}
& S4 kNE subgraph     & 3020  & 6386   & 86.0   & 0.028 & 73 \\
& S4 induced subgraph & 200   & 601    & 11.0   & 0.055 & 73 \\
& S4 SPNSA subgraph   & 306   & 538    & 17.0   & \textbf{0.056} & 73 \\
\midrule
\multirow{3}{*}{IBM-AML (HI-Small)}
& S4 kNE subgraph     & 1152  & 2475   & 104.0  & 0.090 & 7  \\
& S4 induced subgraph & 200   & 538    & 58.0   & \textbf{0.290} & 8  \\
& S4 SPNSA subgraph   & 293   & 453    & 65.0   & 0.222 & 8  \\
\midrule
\multirow{3}{*}{IBM-AML (HI-Medium)}
& S4 kNE subgraph     & 61595 & 123065 & 3162.0 & 0.051 & 14 \\
& S4 induced subgraph & 200   & 890    & 87.0   & \textbf{0.435} & 15 \\
& S4 SPNSA subgraph   & 367   & 899    & 96.0   & 0.262 & 15 \\
\bottomrule
\end{tabular}%
}
\end{table*}

\bibliographystyle{spmpsci} 
\bibliography{refs} 

\end{document}